\documentclass[aps,prl,twocolumn,amssymb,amsfonts,floatfix,superscriptaddress,longbibliography,notitlepage,footinbib]{revtex4-2}
\usepackage[utf8]{inputenc}
\pdfoutput=1
\usepackage{color}
\usepackage{float}
\usepackage{xcolor}
\usepackage{mathtools,amsmath,amsthm,amsfonts,amssymb}
\usepackage{commath}
\usepackage{physics}
\usepackage{bm}
\usepackage[pdftex]{graphicx}
\usepackage[colorlinks=false]{hyperref}
\usepackage{enumitem}
\usepackage{soul}

\begin{document}

\title{Effects of stickiness on the quantum states of strongly chaotic open systems}

\author{Miguel A. Prado Reynoso} 
\affiliation{Department of Physics, University of Connecticut, Storrs, Connecticut 06269, USA.}

\author{Edson M. Signor}
\affiliation{Department of Physics, University of Connecticut, Storrs, Connecticut 06269, USA.}

\author{Sandra D. Prado}
\affiliation{Instituto de F\'isica, Universidade Federal do Rio Grande do Sul, Porto Alegre CP15051, Brazil.}

\author{Lea F. Santos}
\affiliation{Department of Physics, University of Connecticut, Storrs, Connecticut 06269, USA.}

%%%%%%%%%%%%%%%%%%%%%%%%%%%%%%%%%%%%%%%
%    ABSTRACT
%%%%%%%%%%%%%%%%%%%%%%%%%%%%%%%%%%%%%%%

\begin{abstract}
We investigate the effects of classical stickiness (orbits temporarily confined to a region of the chaotic phase space) to the structures of the quantum states of an open system. We consider the standard map of the kicked rotor and verify that regions of stickiness survive in the strong chaotic regime of the closed classical map. By scanning the system's phase space with a leak, we analyze how stickiness affects the degree of localization of the states of the quantum system. We find an excellent correspondence between the classical dwell time and finite-time Lyapunov exponents with the quantum dwell time and Wehrl entropy of the quantum states. Our approach suggests that knowledge of the structure of the classically chaotic trajectories can be used to determine where to place the leak to enhance or decrease the degree of delocalization of the quantum states.
\end{abstract}

\maketitle

%%%%%%%%%%%%%%%%%%%%%%%%%%%%%%%%%%%%%%%
%%%%%%%%%%%%%% INTRODUCTION %%%%%%%%%%%
%%%%%%%%%%%%%%%%%%%%%%%%%%%%%%%%%%%%%%%

There has been increasing interest in understanding the role of chaos in open quantum systems. Two directions often taken by this research are the analysis of dissipative quantum systems coupled to an environment~\cite{Rotter2009,Akemann2019,Hamazaki2019,villasenor2024,Garcia2024ARXIV,SaARXIV} and the study of the quantum counterpart of classical conservative systems, where the phase-space volume is preserved and a leak or hole is introduced ~\cite{Pianigiani1979,Altmann2013}. The present work addresses the latter case of a leaking chaotic system whose dynamics is identical to the closed counterpart until the particle reaches the leak, where it escapes. 

The introduction of an opening to the phase space of a chaotic system was first done to study open billiards~\cite{Pianigiani1979}. Experimentally, there have been studies of the decay rate of ultracold atoms in an optical billiard with a hole in the boundary~\cite{Friedman2001} and of electrons in a two-lead quantum dot with a stadium billiard shape~\cite{Marcus1993}. But models for leaking systems have been used in various practical problems beyond billiards~\cite{Nagler2007}, including the loss of particles in a tokamak~\cite{Portela2008}, chemical reactions~\cite{Dumont1992},
planetary science~\cite{Nagler2004}, and the morphology of the eigenstates of open quantum systems~\cite{Ermann2009}. There has also been studies of stickiness in leaking systems with a mixed phase space~\cite{Altmann2013}. 

Stickiness refers to chaotic orbits that remain in a region of the phase space for a long time before moving to large distances~\cite{Contopoulos2010,Dvorak1999}. The phenomenon was initially found around islands of stability~\cite{Contopoulos1971} and later near the unstable asymptotic curves of unstable periodic orbits~\cite{Contopoulos2010}. Stickiness finds applications in different fields, from the description of anomalous diffusion~\cite{Livorati2018,Borin2023} to the arms of spiral galaxies~\cite{Voglis2006}. It is usually associated with weakly chaotic systems, where the phase space is mixed and regions of chaotic motion coexist with regions of regular motion. When a leak is introduced in the phase space of these systems, the chaotic components that reach the leak are affected~\cite{Altmann2013}.

In this work, we show that stickiness can persist even in a strongly chaotic classical system, and its presence affects the degree of localization of the states of the corresponding quantum system. Our analysis is done for the classical and quantum Chirikov standard map, which is the discrete map of the kicked rotor~\cite{Chirikov1979}, but the approach is general and can be extended to other strongly chaotic systems.

We introduce a leak on the phase space of the classical map and investigate how the dwell (or escape) time of the trajectories depends on the leak's position. The overall degree of chaoticity of the system increases when the leak covers regions of stickiness. This is verified by computing the dwell time of the trajectories and their finite-time Lyapunov exponents (FTLEs) \cite{OttBook}. The latter is used when studying non-stationary flows and transient dynamics \cite{You2021, Stefanski2010, Drotos2021, Prasad1999,Manos2015,Prado2022,Storm2024}. FTLEs provide a local measure of the exponential divergence of phase-space trajectories over a finite time interval, thus allowing for detecting phase-space structures, such as stickiness \cite{Lapeyre2002, Sales2023, Prado2022b, Shadden2005}.

To explore how the position of the leak affects the properties of the quantum system, we analyze the lifetime of the eigenstates of the propagator (which, in leaking systems, are known as resonances), their Husimi functions (distributions of the quantum states in the phase space), and their phase-space entropy (Wehrl entropy). To ensure the orthonormalization of the states, we employ the Schur decomposition~\cite{schomerus2004,Hall2023}. We find excellent correspondence between the phase-space distribution of the classical trajectories and the Husimi functions of the quantum resonances. Quantitatively, the dependence of the FTLE and the classical dwell time on the leak's position is comparable to that of the Wehrl entropy and the dwell time of the quantum resonances, respectively. When the leak covers regions of stickiness, increasing the level of chaoticity of the system and the values of the FTLEs, the entropies grow, indicating that the quantum resonances become more delocalized. Simultaneously, by reducing the effects of stickiness, the classical and quantum dwell times decrease.

While quantum scars~\cite{Heller1984,Pilatowsky2021} (imprints of classical
unstable periodic orbits on quantum states), supersharp resonances~\cite{Novaes2012,Novaes2013}, and the fractal Weyl law~\cite{Shepelyansky2008} have been extensively investigated and scars have recently received  increasing attention in the context of many-body quantum systems~\cite{Turner2018}, the effects of stickiness to the structures of quantum states have hardly been explored. Our study closes this gap and stimulates further analyses of how phase-space structures can impact both closed and open quantum systems.

%%%%%%%%%%%%%%%%% MODEL %%%%%%%%%%%%%%%
{\em Closed classical and quantum models.--} We consider the paradigmatic standard map for which the classical dynamics and the quantization via Floquet theory are well established. This stroboscopic map describes the Poincar\'e section of a rotor that is periodically kicked and moves freely between kicks. It is given by 
\begin{equation}
\begin{split}
q_{n+1} & =  q_n + p_{n}  
\\
p_{n+1} & = p_n -  \frac{K}{2\pi}\sin{(2\pi q_{n+1})}, 
\end{split}
\label{eq:SM}
\end{equation}
where the canonical conjugated coordinates $q_n$ and $p_n$ are taken mod 1. Time is then counted in terms of the number of iterations of the map. The stochastic parameter $K$ in the equation above is chosen as $K = 10$ to ensure strong chaos. At this value of $K$, there are no visible islands of regularity. 

Due to the toroidal phase space $\mathbb{T}^2$, the quantization of the standard map leads to a finite Hilbert space of dimension $N = 2\pi/ \hbar$ with a discretized position basis $q = k/N$ and $k = 1, \dots, N$. In this basis, the time-evolution operator after one iteration is
\begin{eqnarray}
U_{k,k'} \!= \! \frac{1}{\sqrt{N}} \exp{ \! \frac{i\pi}{N}(k-k')^{2} + \frac{iNK}{2\pi} \! \cos{\left( \! \frac{2\pi}{N}k' \! \right)} \!\!}. 
\label{eq_QSM_closed}
\end{eqnarray}

\begin{figure}[t]
    \centering
    \includegraphics[scale=1.0]{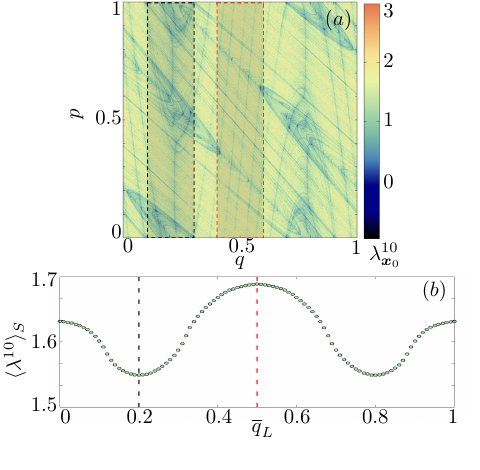}
    \caption{(a) Density plot of the finite-time Lyapunov exponents (FTLEs) for different initial conditions in the phase space of the closed classical standard map in Eq.~(\ref{eq:SM}) after $n=10$ iterations and (b) mean value of the FTLEs for initial conditions in a strip of width $\Delta q=0.2$ centered at $\overline{q}_L$ as a function of $\overline{q}_L$.
    In (a): Two strips are illustrated with shaded areas, one centered at $\overline{q}_L=0.2$ and the other at $\overline{q}_L=0.5$.}
    \label{fig:1}
\end{figure}

%%%%%%%% STICKINESS - CLOSED %%%%%%%%
{\em Stickiness in the closed chaotic classical system.--} In Fig.~\ref{fig:1}(a), we show the density plot of the FTLEs computed after $n=10$ iterations of the closed classical map in Eq.~(\ref{eq:SM}). Despite being strongly chaotic ($K=10$), the phase space of the map exhibits structures associated with regions of stickiness, where the values of the FTLEs are smaller (dark blue) than in the rest of the plot (green to yellow). The shaded rectangular strips centered at $\overline{q}_L=0.2$  and at $\overline{q}_L=0.5$ highlight two very different parts of the phase space, the first marked by regions of stickiness and the latter presenting larger and more homogeneous values of FTLEs. 

To quantitatively identify the regions of stickiness, we compute the average value of the FTLEs  within a strip of length $\Delta q=0.2$ centered at position $x_q$,
\begin{equation}
\langle \lambda^{10}\rangle_S \! = \! \frac{1}{\mu_S} \int_S d\mu \ \lambda_{\boldsymbol{x}_0}^{10},
\end{equation}
where $\boldsymbol{x}_0 = (q_0,p_0)$ denotes the initial condition and  $\mu_S$ represents the measure on the region $S$ of the strip. In Fig.~\ref{fig:1}(b), we show $\langle \lambda^{10}\rangle_S$ as a function of $\overline{q}_L$. In agreement with the structures observed in Fig.~\ref{fig:1}(a), $\langle \lambda^{10}\rangle_S$ is small around $\overline{q}_L=0.2$ (and by symmetry around $\overline{q}_L=0.8$) and large at the vicinity of $\overline{q}_L=0.5$.

%%%%%%%%%% CLASSICAL %%%%%%%%%%%%%%%%
{\em Leaking classical system.--} We now insert a leak in the phase space of the chaotic classical map. The leak corresponds to a strip parallel to the momentum axis, similar to the strips depicted in Fig.~\ref{fig:1}(a). The map recursion remains the same as in the closed system in Eq.~(\ref{eq:SM}) until $(q_n,p_n)$ eventually falls into the hole, when the trajectory is halted. 

The time that the trajectory for a given initial condition remains in the phase space is the classical dwell time, $\tau$. The FTLE associated with that initial condition is determined up to $\tau$ and is denoted by $\lambda_{\boldsymbol{x}_0}^{\tau}$.

 \begin{figure*}[t]
    \centering
    \includegraphics[scale=1.3]{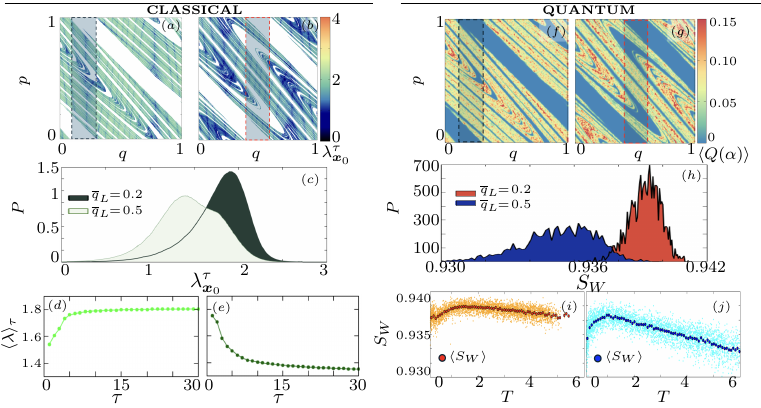}
    \caption{Leaking (a)-(e) classical and (f)-(j) quantum system with a leak at (a,d,f,i) $\overline{q}_L=0.2$ and (b,e,g,j) $\overline{q}_L=0.5$.
    (a)-(b) FTLEs, $\lambda_{\boldsymbol{x}_0}^{\tau}$, for various initial conditions $\boldsymbol{x}_0$. Regions in white correspond to trajectories discarded due to their small dwell times.
    (c) Distributions of the FTLEs in (a) and (b).
    (d)-(e) Average FTLE as a function of the classical dwell time.
    (f)-(g) Husimi functions averaged over 20 Schur states with the largest dwell times to avoid leak-correlated states with fluctuating values of $S_W$. 
    (h) Distributions of the Wehrl entropies $S_W$ of the quantum states.  
    (i)-(j) Wehrl entropies of the Schur vectors as a function of their quantum dwell times. Light circles indicate the values of $S_W$ and dark circles indicate the average, $\langle S_W \rangle$, over intervals of $\Delta T = 0.08$; 
    $N=10^4$ and Husimi's resolution of $10^6$.}
    \label{fig:2}
\end{figure*}

In Figs.~\ref{fig:2}(a)-(b), we show the density plots for the FTLEs of the open classical map, $\lambda_{\boldsymbol{x}_0}^{\tau}$, obtained for various different initial conditions. Dark blue indicates small values of $\lambda_{\boldsymbol{x}_0}^{\tau}$ and green to yellow indicates larger values. White regions correspond to trajectories with small dwell time. We discard these trajectories, because they do not follow the universal exponential decay of the survival probability, that is, for such short times, the ratio between the number of trajectories that remain in the phase space and the total number of initial conditions does not decay exponentially in time~\cite{Altmann2013}. 

Motivated by the results in Fig.~\ref{fig:1}, the leak in Fig.~\ref{fig:2}(a) is centered at $\overline{q}_L=0.2$, so it eliminates a region of the phase-space with predominance of stickiness, while the leak at $\overline{q}_L=0.5$ in Fig.~\ref{fig:2}(b) covers a region with almost no sign of stickiness.  As a consequence, the values of the FTLEs in Fig.~\ref{fig:2}(a) are larger (green to yellow) than the FTLEs (dark blue) in Fig.~\ref{fig:2}(b). The increased degree of chaoticity for the phase space with the leak at the region of stickiness ($\overline{q}_L=0.2$) can also be seen with Fig.~\ref{fig:2}(c), where the distribution of the values of $\lambda_{\boldsymbol{x}_0}^{\tau}$ for this case is right-shifted towards larger values than the distribution of the values of $\lambda_{\boldsymbol{x}_0}^{\tau}$ for  $\overline{q}_L=0.5$.

To provide more detail on how the degree of chaos depends on the position of the leak, we define the average FTLE as
\begin{equation}
\langle \lambda\rangle_{\tau} \! = \! \frac{1}{\mu_\tau} \int_{\Omega_\tau} d\mu \ \lambda_{\boldsymbol{x}_0}^{\tau} ,
\end{equation}
where $\Omega_\tau$ is the set of all initial conditions that give trajectories with dwell time $\tau$ and $\mu_\tau$ represents the natural measure on this set. In Fig.~\ref{fig:2}(d) [Fig.~\ref{fig:2}(e)], we show the average FTLE as a function of $\tau$ for the leak at $\overline{q}_L=0.2$ [$\overline{q}_L=0.5$]. The first trajectories to escape are those close to the leak. When the leak is [not] on the region of stickness, as in Fig.~\ref{fig:2}(d) [Fig.~\ref{fig:2}(e)], those trajectories have low [high] Lyapunov exponent, so $\langle \lambda \rangle_{\tau}$ is small [large] for small $\tau$. The opposite happens for long dwell times, where the surviving trajectories in Fig.~\ref{fig:2}(d) are more chaotic and have larger average FTLE than in  Fig.~\ref{fig:2}(e), even though the total number of surviving trajectories in Fig.~\ref{fig:2}(d) is smaller than Fig.~\ref{fig:2}(e).

%%%%%%%%%% QUANTUM %%%%%%%%%%%%%%%%
{\em Leaking quantum system.--} The propagator $\Tilde{U}$ of the open quantum map is nonunitary. It is obtained by projecting the unitary propagator $U$ of the closed system in Eq.~(\ref{eq_QSM_closed}) on the complement of the leak, $\Tilde{U} = \Pi U$. Since the leak is a strip parallel to the $p$ axis, the projector $\Pi$ is diagonal in the position representation. The eigenvalues of $\Tilde{U}$, denoted by $z_k = \exp{i\theta_{k} - \Gamma_{k}/2}$, contain the quasiangles $\theta_{k}$ and an exponential decaying term, where $\Gamma_{k}$ is the uniform decay rate and $T_{k} = 1/\Gamma_{k}$ is the dwell time of each quantum state. The eigenstates of $\Tilde{U}$, known as resonances, form a nonorthogonal set in which left and right eigenvectors are different, the former being linked to the backward propagation and the latter to the forward propagation of the system. To orthonormalize the states, we use the Schur decomposition~\cite{schomerus2004,Hall2023}.

In Figs.~\ref{fig:2}(f)-(g), we show density plots of the Husimi functions, $Q_k(\alpha) = \exp{-2\pi N|\alpha|^2}|\langle \alpha|v_k \rangle|^2$, which give the distribution of a Schur state $|v_k \rangle$ in the phase space, with $|\alpha\rangle$ being a coherent state centered at $(q,p) \in \mathbb{T}^2$. As in Fig.~\ref{fig:2}(a) [Fig.~\ref{fig:2}(b)], the leak in Fig.~\ref{fig:2}(f) [Fig.~\ref{fig:2}(g)] is at position $\overline{q}_L=0.2$ [$\overline{q}_L=0.5$]. Similarly to the classical analysis in Figs.~\ref{fig:2}(a)-(b), where we consider trajectories with large classical dwell times, Figs.~\ref{fig:2}(f)-(g) show the average of the Husimi functions, $\langle Q(\alpha) \rangle$, over 20 Schur states with the largest quantum dwell times. There is a clear parallel between the classical results in Fig.~\ref{fig:2}(a) [Fig.~\ref{fig:2}(b)] and the quantum results in Fig.~\ref{fig:2}(f) [Fig.~\ref{fig:2}(g)], with regions of smaller values of the FTLEs (dark blue) being reflected by regions of more localization, where the values of $\langle Q(\alpha) \rangle$ are larger  (red).

Analogously to the classical study, where each trajectory is associated with a FTLE, we associate with each Schur state, a measure of delocalization in phase space given by the Wehrl entropy~\cite{Wehrl1978},
\begin{equation}
    S^{(k)}_{W} = - \frac{1}{\mathcal{N}}\int d\alpha \:Q_{k}(\alpha) \ln {Q_{k}(\alpha)} ,
\end{equation}
where  $\mathcal{N}$ is a normalization constant and $0 \leq S_{W} \leq 1$. A fully delocalized state has $S_{W} = 1$ and the uttermost localized state has $S_{W} = 0$. Similarly to the distributions of $\lambda_{\boldsymbol{x}_0}^{\tau}$ in Fig.~\ref{fig:2}(c), we see that the distributions of the Wehrl entropies in Fig.~\ref{fig:2}(h) discriminate the position of the leak. When the leak covers the region of stickiness centered at $\overline{q}_L=0.2$, the distribution of $S_W^{(k)}$ is right-shifted if compared to the distribution of $S_W^{(k)}$ obtained for $\overline{q}_L=0.5$. That is, by eliminating regions of stickiness, the quantum states become more delocalized.

Figures~\ref{fig:2}(i)-(j) display the Wehrl entropies of the Schur states with respect to their dwell times $T$. Light circles correspond to the entropy for a single state, $S_{W}$, and dark circles give the average, $\langle S_{W} \rangle$, over states in an interval $\Delta T = 0.08$. 
In agreement with the classical results in Figs.~\ref{fig:2}(d)-(e), when we get rid of stickiness with the leak at $\overline{q}_L=0.2$ in Fig.~\ref{fig:2}(i), both  $S_{W}$ and $\langle S_{W} \rangle$ reach larger values, indicating a greater degree of  delocalization, than in Fig.~\ref{fig:2}(j), where $\overline{q}_L=0.5$.      

\begin{figure}[t]
        \includegraphics[scale=0.8]{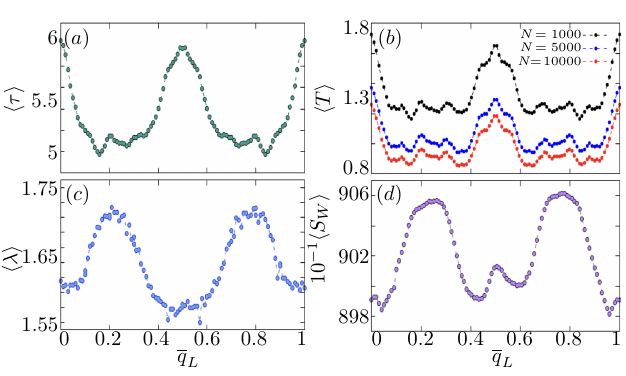}
    \caption{Classical-quantum correspondence and dependence on the position $\overline{q}_L$ of the leak.
    (a) Classical dwell time and (c) FTLE, both averaged over all points of the phase space, as a function of $\overline{q}_L$. 
    (b) Quantum dwell time and (d) Wehrl entropy, both averaged over all $N$ Husimi-Schur vectors, as a function of $\overline{q}_L$.
    In (b): Three different Hilbert space sizes are considered. In (d):  $N=5000$ and Husimi's resolution of $10^6$.}
    \label{fig:3}
\end{figure}

%%%%%%% CLASSICAL-QUANTUM %%%%%%%%%%%%
{\em Classical-quantum correspondence.--}  Figure~\ref{fig:3} provides a general picture of the correlation between stickiness and the properties of leaking systems. Exhibiting excellent classical-quantum correspondence, the average classical dwell time in Fig.~\ref{fig:3}(a) and the average quantum dwell time in Fig.~\ref{fig:3}(b) decrease when the leak removes a region of stickiness, that is, $\langle \tau \rangle$ and $\langle T\rangle$ reach their smallest values for strips placed in the vicinity of  $\overline{q}_L=0.2$ and $\overline{q}_L= 0.8$ (cf. Fig.~\ref{fig:1}). Likewise, the average FTLE in Fig.~\ref{fig:3}(c) and the average Wehrl entropy in Fig.~\ref{fig:3}(d) increase when stickiness is eliminated. In Fig.~\ref{fig:3}(b), we display results for different Hilbert space sizes, $N$. The dependence  on the position of the leak remains the same, and the values of $\langle T \rangle$ converge as $N$ increases.

%%%%%%%%%% CONCLUSIONS %%%%%%%%%%%%
{\em Conclusions.--} Typically, the phase space of hard-chaotic systems does not present structures. Yet, our analysis of the finite-time Lyapunov exponents (FTLEs) of the strongly chaotic classical standard map revealed persistent regions of stickiness. By introducing a leak in the phase space of this system, we were able to analyze the effects of stickiness on both the classical map and its quantum counterpart.

When the leak is placed on a region of stickiness, thus removing the influence of this region, the overall degree of chaoticity of the classical system increases. The FTLEs grow and the classical dwell times decrease. These changes are directly manifested in the quantum system, where the quantum dwell times also decrease and the quantum states become more spread out in phase space, as quantified with the Wehrl entropy. The dwell time and entropy work as quantum probes of stickiness influence. These results imply that by properly choosing where to place a hole in the classical phase space, one can control the level of delocalization of a quantum system. 

Our method to detect and analyze the effects of stickiness can be extended to other strongly chaotic systems. It would also be interesting to investigate open systems with mixed phase spaces, where elaborated structures of stickiness and stability islands should emerge. In short, we have shown that stickiness is one more element worth attention in the current studies of chaos and localization in open systems.

%%%%%%%%%%%%%%%%%%%%%%%%%%%%%%%%%%%%%%%
%    ACKNOWLEDGMENTS
%%%%%%%%%%%%%%%%%%%%%%%%%%%%%%%%%%%%%%%
{\it Acknowledgments.--} This research was supported by the
NSF CCI grant (Award Number 2124511).

%%%%%%%%%%%%%%%%%%%%%%%%%%%%%%%%%%%%%%%
%    BIBLIOGRAPHY
%%%%%%%%%%%%%%%%%%%%%%%%%%%%%%%%%%%%%%%
%\bibliography{biblioNSF2019}

%apsrev4-2.bst 2019-01-14 (MD) hand-edited version of apsrev4-1.bst
%Control: key (0)
%Control: author (8) initials jnrlst
%Control: editor formatted (1) identically to author
%Control: production of article title (0) allowed
%Control: page (0) single
%Control: year (1) truncated
%Control: production of eprint (0) enabled
%

\end{document}